\documentclass{xeus}
\usepackage{graphics}
\usepackage{psfig}

\begin{document}
\onecolumn
\title{X-ray Astronomical Polarimetry in the XEUS Era}
\author{E. Costa\inst{1}, P.Soffitta\inst{1}, G.Di Persio\inst{1},
M.Feroci\inst{1}, L.Pacciani\inst{1}, A.Rubini\inst{1} \and
R.Bellazzini\inst{2}, A.Brez\inst{2}, L.Baldini\inst{2},
L.Latronico\inst{2}, N.Omodei\inst{2}, G.Spandre\inst{2}}
\institute{IASF-CNR,Rome, Italy \and INFN,Sezione di Pisa, Italy}
\authorrunning{E.Costa et al.}
\maketitle

\begin{abstract}

X-ray Polarimetry is almost as old as X-ray Astronomy. Since the
first discovery of X-ray sources theoretical analysis suggested
that a high degree of linear polarization could be expected  due
either to the, extremely non thermal, emission mechanism or to the
transfer of radiation in highly asymmetric systems. The actual
implementation of this subtopic was, conversely, relatively
deceiving. This is mainly due to the limitation of the
conventional techniques based on the Bragg diffraction at
$45^{o}$, or on Thomson scattering around $90^{o}$. Acually no
X-ray Polarimeter has been launched since 25 years. Nevertheless
the expectations from such measurement on several astrophysical
targets including High and Low Mass X-Ray Binaries, isolated
neutron Stars, Galactic and Extragalactic Black Holes is extremely
attractive. We developed a new technique to measure the linear
polarization of X-ray sources. It is based on the visualization of
photoelectron tracks in a, finely subdivided, gas filled detector
(micropattern). The initial direction of the photoelectron is
derived and from the angular distribution of the tracks the amount
and angle of polarization is computed. This technique can find an
optimal exploitation in the focus of XEUS-1. Even in a very
conservative configuration (basically the already existing
prototype) the photoelectric polarimeter could perform polarimetry
at $\%$ level on many AGNs. Further significant improvements can
be expected from a technological development on the detector and
with the use of XEUS-2 telescope.

\end{abstract}

\section{Introduction}
Historically we can group the measurements performed on
Astronomical X-ray Sources into four groups:

\begin{itemize}
\item Timing Photometry (Geiger, Proportional Counters, MCP) with
Rockets, UHURU, Einstein, EXOSAT, ASCA, SAX,  XMM, Chandra.

\item Imaging:

Pseudo-imaging (modulation collimators, coded masks), SAS-3,
XTE-ASM, SAX-WFC, HETE-2

Real Imaging (grazing incidence optics + Position Sensitive
Detectors: IPC, MCA, CCD) with Rockets, Einstein, EXOSAT, ROSAT,
ASCA, SAX, Chandra, Newton.

\item Spectroscopy:

Non dispersive (Proportional Counters, Si/Ge and CCD) Rockets,
Einstein, EXOSAT, HEAO-3, ASCA, SAX, Chandra, Newton.

Dispersive: (Bragg, Gratings) Einstein, Chandra, Newton

\item Polarimetry   (Bragg, Thomson/Compton) with rockets Ariel-5, OSO-8
\end{itemize}

While in the domain of Photometry, Imaging and Spectroscopy the
observing techniques have been tremendously improved, Polarimetry
has only been based on the same, conventional techniques,
producing important but very limited results. In fact, after
OSO-8, no astronomical Polarimeter has been flown any more.

\section{Why X-ray Astrophysical Polarimetry?}

Polarization from celestial sources may derive from:

\begin{itemize}
\item Emission processes themselves: cyclotron, synchrotron,
non-thermal bremmstrahlung (Westfold,
1959; Gnedin $\&$ Sunyaev, 1974; Rees, 1975)

\item Scattering on aspherical accreting plasmas: disks, blobs,
columns (Rees, 1975; Sunyaev $\&$ Titarchuk, 1985;
$M\acute{e}z\acute{a}ros$ et al. 1988).

\item Resonant scattering of lines in hot plasmas (Sazonov 2002).

\item Vacuum polarization and birefringence through extreme magnetic
fields (Gnedin et al., 1978; Ventura, 1979;
$M\acute{e}z\acute{a}ros$ $\&$ Ventura, 1979)
\end{itemize}

\section{Polarization expected in X-ray Pulsators}

The role that polarimetry can play in these sources  is
straightforward. We know that the emission mechanisms and we know,
because we directly measure, the rotation period of the neutron
star (from the light curve) and, in some cases, the intensity of
the magnetic field (from cyclotron lines), and the masses (or at
least the mass ratio) from optical spectroscopy and possibly from
doppler effect on the X-ray period. But other important parameters
such as the inclination of the magnetic to the mechanical axis or
the inclination of the rotation axis on orbit plane are free
parameters to be derived from fitting data of spectral
variability. So far we do not know whether or when the emission is
in the form of a fan or of a pencil. As computed in detail by
$M\acute{e}z\acute{a}ros$ et al. (1988) the polarization of the
cyclotron emission and the different scattering cross section
produces a high degree of linear polarization strongly variable
with energy and phase. With a pencil beam the degree of
polarization will be anticorrelated with the luminosity, while for
a fan beam it will be correlated. We will actually \emph{see} the
magnetic axis swinging around the rotation axis projected on the
sky. All the geometry of the system will be completely frozen.

\section{Polarization expected in isolated Neutron Stars}
Radiation can be polarized when crossing an extreme magnetic field
for the birifringence. Soft thermal X-ray radiation is produced by
a NS atmosphere at $T_{eff}$ of 0.3 to 3 x $10^{6}$ K. The opacity
of a magnetized plasma depends on polarization. While the effects
of magnetic field on the spectrum are negligible the effects on
polarization are outstanding. The degree of polarization
(10$\%$-30 $\%$) depends on photon energy, Teff, magnetic field
and mass-to-radius ratio. Pavlov $\&$ Zavlin, 2000. In radio
pulsars with thermal X-ray emission, phase resolved
polarimetry,will provide mapping the magnetic field. Even more
dramatic effects are expected in Soft Gamma Repeaters, in the
frame of the magnetar model.

\section{Polarization from Scattering in Accretion Disks and General Relativity effects}
Intrinsically unpolarized radiation can be polarized by scattering
as well, provided that scattering angles to the observer are
selected by the system geometry. Chandrasekhar (1960) computed the
maximum polarization (17$\%$)that can derive from an infinitely
extended, infinitely thin scattering cloud. In accretion disks
around compact objects photons are Compton scattered by high
energy electrons have an energy substantially different from
parent population. Therefore in the X ray range the radiatin can
be highly polarized either in the direction perpendicular to the
major axis of the disk or in that parallel (Lightman $\&$ Shapiro,
1976, Sunyaev $\&$ Titarchuk, 1985).

The polarization properties are altered by gravitational effects.
The polarization plane rotates continuously with energy because of
light bending predicted by General Relativistic effects combined
with the radial temperature distribution in the disk. This is a
signature of the presence of a black-hole (Stak$\&$ Connors,
Connors$\&$ Stark, 1977, Connors, Piran $\&$ Stark, 1980).

\section{Polarization of AGNs}
In Seyfert Galaxies and QSOs the effects of scattering, kinematics
and GR are all combined. Moreover the disk/torus geometry produces
significant selection effects in the scattering angles. Also, in
condition of high accretion rate, the X-ray illuminated disk can
be altered in its ionization and temperature structure producing
polarized radiation at energies of 2-6 keV (Matt, Fabian $\&$
Ross, 1995).

Blazar emission will be synchrotron at lower energies and is
expected to be highely polarized (as in IR). At higher energies
inverse compton will prevail and the degree of polarization should
decrease and the angle rotate. From the energy resolved
polarimetry the geometry and energy distribution of the electrons
within the jet can be studied (Poutanen, 1994).

\section{Miscellaneous Targets}
\begin{itemize}
    \item Non thermal X-ray emission from pulsars (for Crab P $<8$$\%$)

    \item Pulsations in LMXB (QPOs and millisecond pulsar)

    \item Edge-on X-ray binaries (polarized lines?)

    \item Jets in Galactic Miniquasars (synchrotron?)

    \item Non thermal components in thermal Supernova Remnants (local or
extended)

    \item Non thermal component in Galaxy Clusters

    \item Gamma-Ray Burst Afterglows

    \item Solar Flares

\end{itemize}

\section{Conventional Techniques}

Compared with great expectations of the theoretical analysis, the
experimental results are quite meagre. In the beginning of X-ray
astronomy polarimeters were flown aboard rockets, and satellites
ARIEL-5 and OSO-7. In practice the only positive result was the
detection of polarization by Crab Nebula by the team of Columbia
University, with a rocket and with OSO-8: 19.2 $\%$ at 2.6 keV
(Weisskopf et al., 1978).

This is mainly due to the limitations of the conventional
techniques of Bragg diffraction and Compton scattering. A Bragg
crystal, operated at $45^{o}$, and rotated around the optical
axis, is an excellent analyzer of Polarization. It preserve
imaging but the efficiency is very poor. Compton around $90^{o}$
is a good compromise of efficiency and modulation, but, unless the
energy lost in the scattering is measured (what is typically
possible at higher energies only), completely destroys positional
information and results in set-up huge, with high background and
very serious systematic effects, that are partially removed by
rotating the whole set-up. The best implementation of these two
techniques is the Stellar X-Ray Polarimeter (Kaaret et al. 1990),
made for the SPECTRUM-X-Gamma Mission, so far not flown.

\section{Photoelectric Polarimetry}

The distribution of electrons in photoelectric effect is good
analyzer of polarization, almost at the same level of Bragg
diffraction, but involves a large slice of the X-Ray spectrum.
When the photon is absorbed by the inner shells of an atom, a
photoelectron is ejected with a kinetic energy which is the
difference of the photon energy and the binding energy of the
electron. The photoelectron is preferentially ejected (actually
with a $cos^{2}$ distribution) on a plane perpendicular to the
incoming photon. Within this plane the ejection directions are
peaked around the electric field of the photon (again with a
$cos^{2}$ distribution).

The photoelectron interacts with the matter around the initial
atom by several processes, two of which are almost exclusively
determining its kinematics: it is slowed by ionizing collisions
with atomic electrons and scattered by coulomb diffusion on the
charge of the nuclei. The photoelectron, just like any electron of
any other origin, leaves in the absorber a track, namely a string
of electron/ion pairs, topologically connected, marking the path
from the creation to the stopping point. All this Physics was
studied in detail by Auger in 1926 by means of a cloud chamber
filled with various mixtures of gas. The tracks of the
photoelectrons, created along the X-ray beam path, are visualized
by the cloud chambers as chains of bright dots. Incidentally by
studying the images Auger discovered the presence of an additional
electron of fixed energy produced by the self-ionization of the
excited ion, since then named the Auger Electron. The cloud
chamber, thanks to the low density of the conversion/detection
material, is an ideal tool to microscopically resolve the
photoelectron track. But a clod chamber is definitely a ground
based device.

Many workers tried in the past to design sensitive x-ray
polarimeters based on the photoelectric effect but with scarce or
no success. Some are based on a combination of a solid
photo-cathode and an electronic detector and they require very
high grazing incidence and pointing stability, while they do not
provide energy information on the X-ray flux.  Other attempt, with
a single integrated analyser and detector, were frustrated.
Actually they detected polarisation only as an 'edge' effect,
either by counting coincidence in neighbors proportional counters
wires (Riegler G.R. et al. Bull. Am. Phys. Soc., 15, 1970, 635.)
or  in neighbors CCD pixels (Tsunemi et al., NIM, 1992).  Only
with the advent of finely segmented gas detectors it now possible
to detect polarization, with the highest sensitivity, in the
canonical energy band for X-ray Astronomy. This approach has been
attempted by means of gas luminescent detectors read with a CCD
through an imaging optics (Austin 1993, La Monaca 1998, Sakurai
2001), but the capability to efficiently apply this method to low
energ X-rays is still to be verified. In the following we present
a newly developed detector, already available for a space
experiment.

\section{A new device: the MICROPATTERN DETECTOR}

Position sensitive gas detectors, such as the Multi-Wire
Proportional Chamber, typically come out with a single information
on  a X-ray event such as the center of gravity or the cross-over
time. This information includes all data on the photoelectron
track. In this sense the  extension of the track is usually
considered as a $\emph{noise}$, something to be kept as small as
possible in the design. For the Multi-Wire Proportional Counter
the extension of the track is considered the ultimate limit to the
space resolution. Our approach is orthogonal. We image the track
to reconstruct the interaction point and the prime direction of
the photoelectron: something very similar to the cloud chamber but
including electronic read-out, measurement of the deposited
energy, self trigger capability, moderate encumbrance .

\begin{figure}
\caption[]{Design of the Micropattern detector with GEM and
readout plane with exagonal pads}
\end{figure}

\begin{figure}
\caption[]{Detector plane with exagonal pads currently working as
laboratory prototype }

\end{figure}

This modern Cloud Chamber is the Micropattern Gas Chamber (Costa,
Bellazzini et al., Nature 2001). It consists (fig. 1, fig.2) of a
gas cell with a drift region, a multiplication stage (actually a
Gas Electron Multiplier) and a multi pixel true bi-dimensional
read-out anode. We have constructed a multi-pixel hexagonal
read-out built on an many-layer PCB (fig.3). The  high granularity
allows the tracks of individual photoelectrons emitted by each
incident X-ray to be followed. The device combines almost the best
performances of gas detectors: pixel sizes from 50 to 200 $\mu$m
are feasible, the signal is very fast(tens of ns), and the energy
resolution reasonable, close to the optimum for such devices
(10$\%$ at 6 keV). Each pixel is connected to a pre-amplifier and
ADC channel which allow to detect the energy lost in that pixel.
The images of the tracks contain therefore also the information of
the dynamics of the photoelectron energy loss and of the energy of
the primary photon. By taking the signal from the GEM we trigger
the acquisition of the anode signal and perform an optimal pulse
height analysis. We collect a track for each detected photon (see
fig 4).

\begin{figure}
\caption[]{Close-up view of the hexagonal pads as readout plane of
the laboratory prototype }

\end{figure}

\begin{figure}
\caption[]{Photoelectron tracks produced in gas in the laboratory
prototype by 5.4 keV photons}
\end{figure}

\begin{figure}
\caption[]{Locations of the Baricentres for 5.4 keV polarized
photons as derived by the tracks detected by the Micropattern}
\end{figure}

The actual track is made as a skein and from its analysis is
always possible to identify the 'head' which carries most of the
information on the polarization from the tail which does not. We
collected tracks from a very finely collimated 5.4 keV unpolarized
sources. The loci of the centroids of each track are located on a
circular region around the interaction points indicating the
tracks, even at this low energy,  are not randomized (fig.5). From
each track we reconstructed the emission angle. The histogram of
the emission angles is indicative of the presence of polarization
in the incoming X-ray photons. In case of non polarised X-ray
photons, such as fluorescence lines,  all the emission angles have
the same probability and the histogram is, therefore, a flat
curve. We measured a flat curve from the fluorescence line
produced by an Fe$^{55}$ source at 5.9 keV or Chromium lines at
5.4 keV indicating that no major spurious effects were present
(fig. 6.1). We, instead measured a significative deviation from a
flat curve when we shined the detector with a polarized X-ray
source of 5.4 keV (fig.6.2).

\begin{figure}
\caption[]{Modulation curve measured with an unpolarized source
$Fe^{55}$ (left panel) and a polarized 5.4 keV source (right
panel) }
\end{figure}

This data are well modelled, also quantitatively, with what we
expect if we take into account the theoretical distribution of the
photoelectron and the smearing effect due to the scattering and to
the lateral diffusion of electrons in the drift from the
absorption point to the GEM.

Therefore we built a detector with combined Polarimetric, Imaging,
Spectral and Timing capabilities and master the simulation tools
to design different configurations dedicated to a particular
experiment set-up.

\section{A  MICROPATTERN detector in the focus of XEUS-1.}

In order to evaluate the capabilities of such a device in the
focus ox XEUS we follow a very conservative approach. We assume to
build a detector with the main features of an existing prototype,
extended to include the whole XEUS PSF. We compute the expected
counting rates from sources and from background (from literature
data) and the modulation factor from monte.carlo simulations
confirmed from our prototype measurements. This can be mounted
within a conventional set-up for gas counters such as the BeppoSAX
MECS GSPC, with a ceramic body and a thin (50$\mu$m) Beryllium
window . This is far from an optimal device but is something of
which we can guarantee, since now the feasibility, and is capable
to perform on a representative sample of celestial objects the
large majority of measurements foreseen in the literature,
including AGNs.

\begin{table}
      \caption{MDP for AGNs in $10^{5}$ s in the 2-10 keV energy band with XEUS-1}
         \label{KapSou}
      \[
         \begin{array}{p{0.5\linewidth}r}
            \hline
            \noalign{\smallskip}
               AGNs     &  {\rm MDP \%} \\
            \noalign{\smallskip}
            \hline
            \noalign{\smallskip}
              CENA                          &  0.6 \\
              NGC4151                            &  0.7 \\
              NGC5548                            &  0.8 \\
              MCG 6-30-15                        &  1.2 \\
              Circinus Galaxy                    &  2.8 \\
              IC4329A                            &  0.7 \\
              Fairall 9                          &  1.6 \\
              MKN501 (Outburst)                  &  0.5 \\
              MKN421                             &  0.7 \\
              3C273                              &  0.9 \\
            \noalign{\smallskip}
             \hline
         \end{array}
      \]
\begin{list}{}{}
\item[$^{\rm a}$] This is a footnote
\end{list}
\end{table}

In Tab.1 we show the Minimum Detectable Polarization for a sample
of bright AGNs with a one day observation with such a device in
the focus of XEUS-1. Energy resolved Polarimetry in 3-4 bands is
possible as well at 1-2 $\%$ level on all of them. Of course a
much more detailed study is possible on Galactic Sources. We
stress the point that in our imaging we can use the reconstructed
absorption point (an not the centroid an in a MWPC). This means
that we have a (experimentally verified) position resolution of
around 100$\mu$m, suitable to exploit all the quality of XEUS
optics. Therefore XEUS could perform angular resolved polarimetry:
e.g. independent polarimetry of an AGN and of its jet; polarimetry
of individual knots of a SNR, polarimetry of regions of a cluster
suspect to host non-thermal components. Also the system wold have
full timing capability and could perform time-resolved
polarimetry: e.g. phase and energy resolved polarimetry of
binaries and of radiopulsars and SGR. A last point we want to
stress is that, as verified at first order with laboratory
testing, systematic effects on this polarimeter are well under
control and no rotation is needed to remove them (as in a
conventional scattering or bragg polarimeter).

\begin{figure}
\caption[]{Effective area of two micropattern detector for
low-energy (0.1-2 keV) and high-energy (2-10 keV) application at
the focus of zero-growth XEUS-1 mirrors. In full colors are
represented the energy bands of X-ray polarimetric sensitivity. }
\end{figure}

\section{Improvements with XEUS-1 and XEUS-2}

The conservative configuration giving the results in the table is
in no way optimal. The efficiency and modulation of the detector
can be improved by increasing the absorption gap and modifying the
gas filling mixture to reduce the scattering and/or the diffusion
on the drift, the two major effects that smear the track. Also the
algorithms so far used are relatively simple and can be improved
by techniques of pattern recognition.

Another important improvement may be achieved by using instead of
one detector, covering the whole energy range, two detectors
optimized in two different bands. The complexity increases but the
total time to achieve a broad band measurement will be
significantly reduced. In fig.7 we show the effective area of our
conservative configuration (on the right), effective from 2 to 10
keV, and (on the left) of a low energy detector, optimized in the
band 0.6 - 3 keV. Such a low energy device could allow for 1$\%$
polarimetry of MK421 in $10^{4}$s.

A further important improvement can derive from the implementation
of XEUS-2 optics. From statistics the MDP will scale with the
square root of areas. Even though we cannot nowadays be sure of
our capability to control systematic effects to perform reliable
polarimetry below 1$\%$, the observations will be significantly
shorter and the sample will become much richer including AGNs at
higher red-shift.

\section{Conclusions}

We conclude that with the new MICROPATTERN device, the Polarimetry
of Astrophysical sources is now feasible, provided that a high
throughput optics is used. This will open a new window in the sky
and dramatically improve our understanding of Physics of X-ray
emitting regions around NS and Black Holes. With XEUS-1 optics,
and with moderate assumptions on technological developments,
polarimetry to the $\%$ level of tens of AGNs will be feasible.

Therefore we think that the inclusion of a MICROPATTERN
photoelectric  polarimeter in the baseline payload for XEUS-1
should be seriously considered.


\end{document}